\documentstyle[12pt,epsfig]{article}
\newcommand{\n}{\hspace*{-2.5mm}}
\newcommand{\arcosh}{\mathop{\rm arcosh}\nolimits}

\catcode`@=11
\newcount\@tempcntc
\def\@citex[#1]#2{\if@filesw\immediate\write\@auxout{\string\citation{#2}}\fi
  \@tempcnta\z@\@tempcntb\m@ne\def\@citea{}\@cite{\@for\@citeb:=#2\do
    {\@ifundefined
       {b@\@citeb}{\@citeo\@tempcntb\m@ne\@citea\def\@citea{,}{\bf ?}\@warning
       {Citation `\@citeb' on page \thepage \space undefined}}%
    {\setbox\z@\hbox{\global\@tempcntc0\csname b@\@citeb\endcsname\relax}%
     \ifnum\@tempcntc=\z@ \@citeo\@tempcntb\m@ne
       \@citea\def\@citea{,}\hbox{\csname b@\@citeb\endcsname}%
     \else
      \advance\@tempcntb\@ne
      \ifnum\@tempcntb=\@tempcntc
      \else\advance\@tempcntb\m@ne\@citeo
      \@tempcnta\@tempcntc\@tempcntb\@tempcntc\fi\fi}}\@citeo}{#1}}
\def\@citeo{\ifnum\@tempcnta>\@tempcntb\else\@citea\def\@citea{,}%
  \ifnum\@tempcnta=\@tempcntb\the\@tempcnta\else
   {\advance\@tempcnta\@ne\ifnum\@tempcnta=\@tempcntb \else \def\@citea{--}\fi
    \advance\@tempcnta\m@ne\the\@tempcnta\@citea\the\@tempcntb}\fi\fi}
\catcode`@=12

\begin{document}
\title{\vskip-3cm{\baselineskip14pt
\centerline{\normalsize\hfill FERMILAB--PUB--95/247--T}
\centerline{\normalsize\hfill MPI/PhT/95--74}
\centerline{\normalsize\hfill hep-ph/9403386}
\centerline{\normalsize\hfill July 1995}
}
\vskip1.5cm
Dependence of Electroweak Parameters on the Definition of the Top-Quark Mass}
\author{{\sc Bernd A. Kniehl}\thanks{Permanent address:
Max-Planck-Institut f\"ur Physik, Werner-Heisenberg-Institut,
F\"ohringer Ring 6, 80805 Munich, Germany.}\\
{\normalsize Theoretical Physics Department, Fermi National Accelerator
Laboratory,}\\
{\normalsize P.O. Box 500, Batavia, IL 60510, USA}}
\date{}
\maketitle
\begin{abstract}
The QCD corrections to electroweak parameters depend on the renormalization
scheme and scales used to define the top-quark mass.
We analyze these dependences for the $W$-boson mass predicted via $\Delta r$ to
${\cal O}(\alpha\alpha_s)$ and ${\cal O}(\alpha\alpha_s^2)$ in the on-shell
and $\overline{\mbox{MS}}$ schemes.
These variations provide us with a hint on the magnitude of the unknown
higher-order QCD effects and contribute to the theoretical error of the
prediction.

\medskip
\noindent
PACS numbers: 12.15.--y, 11.15.Me, 12.15.Lk, 14.65.Ha
\end{abstract}

\section{Introduction}

It is well known that strong-interactions effects on the vacuum-polarization
functions of the electroweak gauge bosons play a significant r\^ole in present
and future high-precision tests of the standard model (SM)
\cite{hal,sto,fan,alt}.
In perturbative calculations to ${\cal O}(\alpha\alpha_s)$
\cite{cgn,ver,kni,djo}, these effects arise from the type of two-loop diagrams
where one virtual gluon is exchanged within a quark loop inserted into an
electroweak-gauge-boson line.
Since the top quark is by far the heaviest established elementary particle,
with a pole mass of $M_t=(180\pm12)$~GeV \cite{abe}, the leading high-$M_t$
terms, of ${\cal O}(G_FM_t^2)$, are particularly important.
As for oblique corrections, {\it i.e.}, those which arise from the gauge-boson
vacuum polarizations, these terms together with their quantum-chromodynamical
(QCD) corrections are all concentrated in $\Delta\rho=1-1/\rho$, where $\rho$
is the familiar parameter introduced in Ref.~\cite{vel}.
The leading-order QCD corrections to $\Delta\rho$ have been known for several
years \cite{ver,kni};
those of next-to-leading order have recently been calculated \cite{avd} and
found to be relatively large, having indeed a non-negligible impact on ongoing
precision tests of the SM.
It is of great phenomenological interest to estimate the residual theoretical
uncertainty of the QCD corrections to $\Delta\rho$ and other basic electroweak
parameters such as $\Delta r$ \cite{sir}.
This is the motivation for the present paper.

QCD-improved analyses of electroweak parameters
\cite{hal,sto,fan,alt,cgn,ver,kni,djo} are usually carried out using the pole
definition of the top-quark mass.
This directly corresponds to the mass parameter which is presently being
extracted with the Fermilab Tevatron \cite{abe} and will be with the CERN
Large Hadron Collider (LHC) and future $e^+e^-$ linear colliders, since,
in the propagation of the $t$ and $\bar t$ quarks between their production and
decay vertices, configurations near their mass shells are kinematically
favoured.
As a matter of principle, however, this mass convention is arbitrary, and we
might as well adopt another one.
For, if all orders of the perturbation expansion were taken into account, the
final result should not depend on the selected scheme.
Yet, this no longer holds true if the perturbation series is truncated.
In general, the finite-order results also depend on the renormalization scales
of the quark masses.
It is generally believed that the scheme and typical scale variations may be
used to estimate the theoretical uncertainty due to the unknown higher-order
corrections.

In this paper, we shall pay special attention to $\Delta r$ \cite{sir}, which
parameterizes the non-photonic corrections to the muon lifetime and allows us
to indirectly determine the $W$-boson mass for given values of the top-quark
and Higgs-boson masses.
Using two popular definitions of quark mass in QCD, namely the pole mass and
the mass of the modified minimal-subtraction ($\overline{\mbox{MS}}$) scheme
\cite{msb}, we shall quantitatively analyze the scheme and scale dependences
of the $M_W$ prediction at next-to-leading order in QCD, and so estimate the
residual theoretical error on $M_W$ from QCD sources.
Our evaluation of $\Delta r$ will make full use of the present knowledge of
higher-order corrections, so that the extracted $M_W$ values should be reliable
within the quoted errors.
Confrontation of these values with the future high-precision measurements of
$M_W$ at the Tevatron and the CERN Large Electron-Positron Collider (LEP2),
in connection with a reduced error on $M_t$, will allow us to pin down the mass
of the SM Higgs boson and to facilitate the search for it.

At this point, a few comments on the so-called $t\bar t$ threshold effects
\cite{thr,ynd} are in order.
The study of these effects was an attempt to estimate the dominant higher-order
QCD corrections to $\Delta\rho$ and other oblique electroweak parameters prior
to their diagrammatical computation \cite{avd}.
This approach was based on the assumption that the bulk of the QCD corrections
arises from the ladder diagrams of multi-gluon exchange in the $t\bar t$
system.
The absorptive parts of these diagrams can be resummed in the non-relativistic
approximation, producing a prominent enhancement of the $t\bar t$ excitation
curve in the threshold region along with a lowering of its onset.
This treatment naturally takes into account the finite lifetime of the top
quark as well.
The real parts of the diagrams were then found via dispersion relations with
subtractions derived from Ward identities.
The fact that the explicit ${\cal O}(\alpha_s^2G_FM_t^2)$ calculation of
$\Delta\rho$ \cite{avd} nicely agrees with the rough $t\bar t$-threshold
estimate \cite{thr}---in fact, it comfortably lies within the errors quoted in
Ref.~\cite{thr}---may be viewed as some posterior justification of this method.
However, the good agreement is, to some extent, fortuitous, since
approximately 30\% of the ${\cal O}(\alpha_s^2G_FM_t^2)$ correction is due to
double-triangle diagrams \cite{avd}, which are beyond the scope of
Ref.~\cite{thr}.
Therefore and for other reasons given in Ref.~\cite{yr}, we shall take the
point of view that, at the present time, the result of Ref.~\cite{avd}
represents the most reliable description of the QCD corrections to $\Delta\rho$
beyond the leading order, and that the residual theoretical uncertainty may be
assessed by analyzing scheme and scale variations.

This paper is organized as follows.
In Section~2, we shall translate existing on-shell (OS) results for the
gauge-boson vacuum polarizations in ${\cal O}(\alpha\alpha_s)$ \cite{kni} to
the $\overline{\mbox{MS}}$ scheme of quark-mass renormalization.
Furthermore, we shall take a look inside the renormalization-group (RG)
structure of the QCD expansion of $\Delta\rho$ to
${\cal O}(\alpha_s^2G_FM_t^2)$ \cite{avd}.
In Section~3, we shall quantitatively analyze the scheme and scale dependences
of the $M_W$ value predicted from the analysis of $\Delta r$ to leading and
next-to-leading order in QCD.
Our conclusions concerning the theoretical uncertainty in $M_W$ of QCD origin
are summarized in Section~4.
The Appendix contains some general relations which may be used to implement the
scale dependence of the QCD coupling and mass in the $\overline{\mbox{MS}}$
scheme, and to switch between the $\overline{\mbox{MS}}$ and OS schemes.

\section{Formalism}

We shall work in the electroweak OS renormalization scheme, which uses the
fine-structure constant, $\alpha$, and the physical particle masses as basic
parameters, and define $c_w^2=1-s_w^2=M_W^2/M_Z^2$ \cite{sir,aok}.
Our analysis of $\Delta r$ will be based on Refs.~\cite{hal,sto,fan}.
Following Refs.~\cite{sto,fan}, we shall use the decomposition
\begin{equation}
\label{dr}
\Delta r=\Delta\alpha-{c_w^2\over s_w^2}\Delta\rho(1-\Delta\alpha)
+\Delta r_{rem},
\end{equation}
where $\Delta\alpha$ embodies the contributions from the charged leptons and
the first five quark flavours that drive the fine-structure constant from the
Thomson limit to the $Z$-boson scale, $\Delta\rho$ is the top-quark-induced
shift in the $\rho$ parameter \cite{vel} written with the Fermi constant,
$G_F$, and $\Delta r_{rem}$ is devoid of
large logarithmic and power-like terms of fermionic origin.
We would like to point out that, unlike $\Delta\rho$, $\Delta r_{rem}$ must be
written in terms of $\alpha$, since it is via $\Delta r$ that $G_F$ is
introduced into the SM \cite{hal,sto,fan,sir}.
To our knowledge, all existing analyses to ${\cal O}(\alpha\alpha_s)$ of
$\Delta r$ and other oblique electroweak parameters with a non-trivial
quark-mass dependence, {\it i.e.}, other than $\Delta\rho$, employ the pole
definition of quark mass in QCD \cite{hal,sto,fan,alt,cgn,ver,kni,djo}.
In the following, we shall describe how these calculations may be converted to
other schemes of mass renormalization in QCD, in particular to the
$\overline{\mbox{MS}}$ scheme.

The relevant quantities in this context are the transverse gauge-boson vacuum
polarizations induced by a pair of quarks, with pole masses $M_1$ and $M_2$.
Adopting the conventions of Ref.~\cite{kni}, we may write the vector and
axial-vector parts as
\begin{equation}
\label{dec}
\Pi^{V,A}(s,M_1,M_2)={N_c\over3}\Pi_0^{V,A}(s,M_1,M_2)
+a{N_cC_F\over4}\Pi_1^{V,A}(s,M_1,M_2)+{\cal O}(a^2),
\end{equation}
where $s$ is the square of the external four-momentum, $N_c=3$,
$C_F=\left(N_c^2-1\right)/(2N_c)=4/3$, and $a=\alpha_s/\pi$.
We shall employ dimensional regularization in $n=4-2\epsilon$ space-time
dimensions and introduce a 't~Hooft mass, $\mu$, to keep the coupling constants
dimensionless.
We shall suppress terms containing $\gamma_E-\ln(4\pi)$, where $\gamma_E$ is
Euler's constant.
These terms may be retrieved by substituting
$\mu^2\to4\pi e^{-\gamma_E}\mu^2$.
In the $\overline{\mbox{MS}}$ scheme \cite{msb}, they are subtracted along
with the poles in $\epsilon$.
For later use, we list the lowest-order functions appearing in Eq.~(\ref{dec})
\cite{kks}:
\begin{eqnarray}
\label{lo}
&\n\n&4\pi^2\Pi_0^{V,A}(s,M_1,M_2)=
\left({1\over\epsilon}+\ln{\mu^2\over M_1M_2}\right)
\left[s-{3\over2}(M_1\mp M_2)^2\right]
\nonumber\\
&\n\n&+\left[1+{1\over s}\left(-{M_1^2+M_2^2\over2}\pm3M_1M_2\right)
-{\left(M_1^2-M_2^2\right)^2\over2s^2}\right]
\sqrt\lambda\arcosh{M_1^2+M_2^2-s\over2M_1M_2}
\nonumber\\
&\n\n&+{M_1^2-M_2^2\over2s}\left[{\left(M_1^2-M_2^2\right)^2\over2s}
\mp3M_1M_2\right]\ln{M_1^2\over M_2^2}
+{5\over3}s-M_1^2-M_2^2\pm6M_1M_2
\nonumber\\
&\n\n&-{\left(M_1^2-M_2^2\right)^2\over2s}+{\cal O}(\epsilon),
\end{eqnarray}
where $\lambda=\left[s-(M_1+M_2)^2\right]\left[s-(M_1-M_2)^2\right]$.
Equation~(\ref{lo}) is valid for $s<(M_1-M_2)^2$.
It may be analytically continued to other values of $s$ by observing that
$s$ comes with an infinitesimal, positive imaginary part, {\it i.e.},
$s\to s+i\varepsilon$.
Specifically,
\begin{eqnarray}
\sqrt\lambda\arcosh{M_1^2+M_2^2-s\over2M_1M_2}
&\n=\n&-\sqrt{-\lambda}\arccos{M_1^2+M_2^2-s\over2M_1M_2}\nonumber\\
&\n=\n&-\sqrt\lambda\left(\arcosh{s-M_1^2-M_2^2\over2M_1M_2}-i\pi\right).
\end{eqnarray}
The general formulae for $\Pi_1^{V,A}(s,M_1,M_2)$ are somewhat lengthy
\cite{djo}.
Fortunately, the finite-mass effects in ${\cal O}(\alpha\alpha_s)$ on the
oblique electroweak parameters (with the well-known exception of
$\Delta\alpha$) due to the first five quark flavours are exceedingly small
\cite{kni,djo}, so that, beyond ${\cal O}(\alpha)$, we may safely neglect all
quark masses, except for $M_t$.
In this approximation, the mass-dependent cases may be cast into the form
\cite{kni}
\begin{eqnarray}
\label{twoloop}
{\pi^2\over M^2}\Pi_1^V(s,M,M)&\n=\n&RX_1+V_1(R)+{\cal O}(\epsilon),
\nonumber\\
{\pi^2\over M^2}\Pi_1^A(s,M,M)&\n=\n&RX_1+Y_1+A_1(R)+{\cal O}(\epsilon),
\nonumber\\
{\pi^2\over M^2}\Pi_1^{V,A}(s,M,0)&\n=\n&{1\over4}(XX_1+Y_1)+F_1(X)
+{\cal O}(\epsilon),
\end{eqnarray}
where $V_1$, $A_1$, and $F_1$ are finite functions of $R=(s/4M^2)$ and
$X=s/M^2$, and \cite{djo,hll}
\begin{eqnarray}
\label{uvdiv}
X_1&\n=\n&{1\over2\epsilon}+L-4\zeta(3)+{55\over12},
\nonumber\\
Y_1&\n=\n&{3\over2\epsilon^2}+{1\over\epsilon}\left(3L+{11\over4}\right)
+3L^2+{11\over2}L+6\zeta(3)+{9\over2}\zeta(2)-{11\over8},
\end{eqnarray}
with $L=\ln(\mu^2/M^2)$.
Here, $\zeta$ is Riemann's zeta function, with values $\zeta(2)=\pi^2/6$ and
$\zeta(3)\approx1.202\,057$.

The quantities in Eq.~(\ref{twoloop}) refer to the OS scheme, {\it i.e.}, they
contain the contributions which emerge from the respective one-loop seed
diagrams by inserting the OS mass counterterm,
$\delta M=m_0-M$, where $m_0$ is the bare quark mass, in all possible ways.
The resulting shifts in $\Pi_1^{V,A}$ may conveniently be constructed from
$\Pi_0^{V,A}$ through the variation
\begin{equation}
a{N_cC_F\over4}\delta\Pi_1^{V,A}(s,M_1,M_2)
={N_c\over3}\sum_{i=1}^2\delta M_i{\partial\over\partial M_i}
\Pi_0^{V,A}(s,M_1,M_2).
\end{equation}
For later use, we list the OS mass counterterm \cite{tar}:
\begin{eqnarray}
\label{osct}
{\delta M\over M}&\n=\n&-aC_F\left({\mu^2e^{\gamma_E}\over M^2}\right)^\epsilon
\Gamma(1+\epsilon){3-2\epsilon\over4\epsilon(1-2\epsilon)}+{\cal O}(a^2)
\nonumber\\
&\n=\n&-aC_F\left[
{3\over4\epsilon}+{3\over4}L+1+\epsilon\left({3\over8}L^2+L
+{3\over8}\zeta(2)+2\right)+{\cal O}(\epsilon^2)\right]+{\cal O}(a^2),
\end{eqnarray}
where $\Gamma$ is Euler's Gamma function and $L$ is defined below
Eq.~(\ref{uvdiv}).

In other mass renormalization schemes, $m_0$ is differently split into the
renormalized mass, $m$, and the counterterm, $\delta m$, {\it i.e.},
\begin{equation}
\label{split}
m_0=M+\delta M=m+\delta m,
\end{equation}
where $\delta m$ differs from $\delta M$ by some finite amount.
In general, $m$ depends on the renormalization scale, $\mu$, and on the QCD
gauge parameter.
Of all possible schemes, the $\overline{\mbox{MS}}$ scheme is singled out
because there $\delta m$ just collects the $\epsilon$ poles in
Eq.~(\ref{osct}).
Furthermore, the $\overline{\mbox{MS}}$ mass is gauge independent, as is $M$
\cite{tar}.
We may translate Eq.~(\ref{dec}) to some other scheme by replacing $M_i$ with
$m_i$ and adjusting the mass counterterms.
The latter generates the shift
\begin{equation}
\label{shift}
a{N_cC_F\over4}\Delta\Pi_1^{V,A}(s,m_1,m_2)
={N_c\over3}\sum_{i=1}^2(M_i-m_i){\partial\over\partial m_i}
\Pi_0^{V,A}(s,m_1,m_2),
\end{equation}
where we have used Eq.~(\ref{split}).
Notice that Eq.~(\ref{shift}) does not require knowledge of the
${\cal O}(\epsilon)$ terms of $\Pi_0^{V,A}$.

In the following, we shall take $m$ to be the $\overline{\mbox{MS}}$ mass.
 From Eq.~(\ref{osct}), we may read off that
\begin{equation}
\label{osmslo}
M=m\left[1+aC_F\left({3\over4}l+1\right)+{\cal O}(a^2)\right],
\end{equation}
where $l=\ln(\mu^2/m^2)$.
The ${\cal O}(a^2)$ term of Eq.~(\ref{osmslo}) is written out in
Eq.~(\ref{osms}).
With the help of Eqs.~(\ref{lo}), (\ref{shift}), and (\ref{osmslo}),
we may now convert Eq.~(\ref{twoloop}) to the $\overline{\mbox{MS}}$ scheme.
The resulting expressions emerge from Eq.~(\ref{twoloop}) by replacing
$M$, $R$, $X$, $X_1$, $Y_1$, $V_1(R)$, $A_1(R)$, and $F_1(X)$ with
$m$, $r=(s/4m^2)$, $x=s/m^2$,
\begin{eqnarray}
\label{twoms}
\overline X_1&\n=\n&{1\over2\epsilon}+l-4\zeta(3)+{55\over12},
\nonumber\\
\overline Y_1&\n=\n&{3\over2\epsilon^2}-{5\over4\epsilon}-{3\over2}l^2
-{5\over2}l+6\zeta(3)+3\zeta(2)-{75\over8},
\nonumber\\
\overline V_1(r)&\n=\n&V_1(r)
+(3l+4)\left(-{\arcsin\sqrt r\over r\sqrt{1/r-1}}+1\right),
\nonumber\\
\overline A_1(r)&\n=\n&A_1(r)
+(3l+4)\left(2\sqrt{{1\over r}-1}\arcsin\sqrt r-1\right),
\nonumber\\
\overline F_1(x)&\n=\n&F_1(x)
+{1\over4}(3l+4)\left[\left(1-{1\over x^2}\right)\ln(1-x)-{1\over x}\right],
\end{eqnarray}
respectively, where $l$ is defined below Eq.~(\ref{osmslo}).
The last three lines of Eq.~(\ref{twoms}) are valid for $0\le r,x\le1$,
the range relevant for the evaluation of electroweak parameters.
Expressions appropriate for other values of $r$ and $x$ may be obtained
with ease by analytic continuation as described in Ref.~\cite{sin}.
The following special cases frequently occur in applications:
\begin{eqnarray}
\overline V_1(0)&\n=\n&V_1(0)=0,
\nonumber\\
\overline V_1^\prime(0)&\n=\n&V_1^\prime(0)-{2\over3}(3l+4),
\nonumber\\
\overline A_1(0)&\n=\n&A_1(0)+3l+4,
\nonumber\\
\overline F_1(0)&\n=\n&F_1(0)+{1\over8}(3l+4).
\end{eqnarray}

In the remainder of this section, we shall discuss the incorporation of
three-loop QCD corrections \cite{avd,che} in Eq.~(\ref{dr}).
In Ref.~\cite{avd}, the ${\cal O}(\alpha_s^2G_FM_t^2)$ term of $\Delta\rho$
has been calculated using both the OS and $\overline{\rm MS}$ definitions of
quark mass in QCD.
In Ref.~\cite{che}, the ${\cal O}(\alpha\alpha_s^2)$ correction to
$\Delta r_{rem}$ has been expanded in powers of $M_Z^2/M_t^2$, and the first
two terms of this expansion have been presented both in the OS and
$\overline{\rm MS}$ schemes.
It is instructive to rewrite the results of Ref.~\cite{avd} in such a way that
the RG origin of the logarithmic terms as well as the interrelation between the
OS and $\overline{\rm MS}$ versions are explicitly displayed.
This may be achieved with the help of Eqs.~(\ref{expa}) and (\ref{osms}) for
$n_f=6$ active quark flavours and leads to
\begin{eqnarray}
\label{rhoos}
\Delta\rho&\n=\n&N_cX_t\left[1+aR_1(1+a\beta_0L)+a^2R_2+{\cal O}(a^3)\right]
\nonumber\\
&\n\approx\n&3X_t\left[1-2.859\,912\,a\left(1+{7\over4}aL\right)
-14.594\,028\,a^2+{\cal O}(a^3)\right],\\
\label{rhomsb}
\Delta\bar\rho&\n=\n&N_cx_t\left\{1+a\left[2\gamma_0l
+r_1\left(1\rule{0mm}{4mm}+a(\beta_0+2\gamma_0)l\right)\right]
+a^2\left[\gamma_0(\beta_0+2\gamma_0)l^2+2(-2\gamma_0^2+\gamma_1)l+r_2\right]
\right.\nonumber\\
&\n\n&\left.\rule{0mm}{4mm}+{\cal O}(a^3)\right\}
\nonumber\\
&\n\approx\n&3x_t\left\{1+a\left[2l-0.193\,245\left(1+{15\over4}al\right)
\right]+a^2\left({15\over4}l^2+{11\over4}l-3.969\,560\right)
+{\cal O}(a^3)\right\},\nonumber\\
&\n\n&
\end{eqnarray}
with
\begin{eqnarray}
r_1&\n=\n&R_1+2C_F,\nonumber\\
r_2&\n=\n&R_2+C_F(2R_1-4\gamma_0+C_F)+2K_0,
\end{eqnarray}
where $X_t=(G_FM_t^2/8\pi^2\sqrt2)$, $x_t=(G_Fm_t^2/8\pi^2\sqrt2)$,
and $L$ and $l$ are defined below Eqs.~(\ref{uvdiv}) and (\ref{osmslo}),
respectively.
The first three coefficients of the beta function and the quark-mass anomalous
dimension of QCD are listed in Eqs.~(\ref{beta}) and (\ref{ano}), respectively,
and $K_0$ is defined in Eq.~(\ref{kzero}).
The genuine information on the QCD correction to $\Delta\rho$ is carried by
$R_1$ \cite{ver,kni} and $R_2$ \cite{avd} or, equivalently, by $r_1$
\cite{jeg} and $r_2$ \cite{avd}.

In the OS scheme, all $\mu$ dependence originates in the renormalization of
$a$, while the $\overline{\mbox{MS}}$ scheme has $m$ as an additional source of
$\mu$ dependence.
So far, we have used a common renormalization scale, $\mu$, for $a$ and $m$.
In the following, we shall abandon this restriction and distinguish between
coupling and mass renormalization scales, $\mu_c$ and $\mu_m$.
We shall henceforth use the symbol $\mu_c$ in the OS expressions.
In order to disentangle $\mu_c$ and $\mu_m$ in the $\overline{\mbox{MS}}$
formulae, we first replace $\mu$ with $\mu_m$ and then substitute
\begin{equation}
a(\mu_m)=a(\mu_c)\left[1-a(\mu_c)\beta_0\ln{\mu_m^2\over\mu_c^2}+{\cal O}(a^2)
\right],
\end{equation}
which follows from Eq.~(\ref{expa}).
Finally, we expand the resulting expressions in powers of $a(\mu_c)$ and
truncate these expansions beyond the order under consideration.

In the next section, we shall estimate the QCD-related theoretical uncertainty
in $M_W$ by studying the scheme and scale dependences of the $\Delta r$
analysis.
In order for our estimate to be meaningful, we shall have to judiciously choose
the central values and widths of the scale intervals.
It is natural to define the central values of $\mu_c$ and $\mu_m$ in such a way
that there the radiative corrections are devoid of logarithmic terms.
Looking at Eqs.~(\ref{rhoos}) and (\ref{rhomsb}), we are thus led to set
$\mu_c=\xi_cM_t$ in the OS scheme, and $\mu_c=\xi_c\bar\mu_t$ and
$\mu_m=\xi_m\bar\mu_t$, where $\bar\mu_t=m_t(\bar\mu_t)$, in the
$\overline{\mbox{MS}}$ scheme.
Here, $\xi_c$ and $\xi_m$ are variable numbers of order unity.
A closed expression for $\bar\mu_t$ in terms of $M_t$ may be found in
Eq.~(\ref{muos}).

\section{Numerical analysis}

We are now in a position to quantitatively explore the scheme and scale
dependences of the $M_W$ prediction on the basis of $\Delta r$.
Our starting point is the OS analysis of Ref.~\cite{sto}, which includes the
dominant corrections beyond one loop, of ${\cal O}(\alpha\alpha_s)$ \cite{kni},
${\cal O}(G_F^2M_t^4)$ \cite{bar}, and ${\cal O}(G_F^2M_W^2M_H^2)$
\cite{hal,sto,rba}.
Here, we extend this analysis to ${\cal O}(\alpha\alpha_s^2)$ by accommodating
the respective corrections to $\Delta\rho$ \cite{avd} and $\Delta r_{rem}$
\cite{che} in Eq.~(\ref{dr}).
Furthermore, we update our input parameters according to
Refs.~\cite{pdg,lep,eid,bet}.
In particular, we use the combined LEP1 value $M_Z=91.1887$~GeV \cite{lep},
the value $\Delta\alpha_{had}^{(5)}=0.0280$ \cite{eid} for the hadronic
contribution to $\Delta\alpha$, and the world average
$\alpha_s^{(5)}(M_Z)=0.118$ \cite{bet}.
For the purpose of studying the scheme and scale dependences of $\Delta r$, we
may assume that $M_H$, $M_t$, and $\alpha_s^{(6)}(M_t)$ are precisely known.
Unless stated otherwise, we choose $M_H=300$~GeV \cite{lep} and $M_t=180$~GeV
\cite{abe}.
We compute $\alpha_s^{(6)}(M_t)$ at three loops in two steps.
First, we scale $\alpha_s^{(5)}(\mu_c)$ from $\mu_c=M_Z$ to $\mu_c=M_t$ via
Eq.~(\ref{resa}) with $n_f=5$.
Then, we cross the flavour threshold at $\mu_c=M_t$ using the matching
condition (\ref{mat}).
For $M_t=180$~GeV, we so obtain $\alpha_s^{(6)}(M_t)=0.1071$.
In order for our findings concerning the scheme and scale dependences to be
meaningful, it is crucial that we consistently evaluate $\alpha_s^{(6)}(\mu)$,
$m_t(\mu)$, and $\bar\mu_t$ to the order that we consider at a time.
For the analysis of $\Delta r$ to ${\cal O}(\alpha\alpha_s^2)$, we first
compute $\alpha_s^{(6)}(\mu)$ using Eq.~(\ref{resa}) for $n_f=6$ with the
${\cal O}(\alpha_s^3)$ terms in the denominator omitted.
Then, we insert this value together with $\alpha_s^{(6)}(M_t)$ and $m_t(M_t)$,
which we extract from Eq.~(\ref{pole}), into Eq.~(\ref{twom}) to obtain
$m_t(\mu)$.
We evaluate $\bar\mu_t$ by means of Eq.~(\ref{muos}).
If we calculate $\Delta r$ to ${\cal O}(\alpha\alpha_s)$, then we
correspondingly discard the ${\cal O}(\alpha_s^2)$ terms in Eqs.~(\ref{resa}),
(\ref{pole}), and (\ref{muos}) and employ Eq.~(\ref{onem}) instead of
Eq.~(\ref{twom}).

\begin{table}[ht]
\caption{
$M_W$ predicted for various values of $M_t$ and $M_H$ via $\Delta r$ to
${\cal O}(\alpha\alpha_s)$ and ${\cal O}(\alpha\alpha_s^2)$ in the OS and
$\overline{\rm MS}$ schemes with $\xi_c=\xi_m=1$.
For reference, also the $M_W$ values evaluated without perturbative QCD
corrections are given.
Also $\bar\mu_t$ is listed.
All masses are given in GeV.
}\label{tab:scheme}
\medskip
\begin{tabular}{|c|c|c|r|c|c|c|c|c|} \hline\hline
\rule{0mm}{5mm}$M_t$ & \multicolumn{2}{c|}{$\bar\mu_t$} & $M_H$ &
\multicolumn{5}{c|}{$M_W$} \\ \cline{2-3} \cline{5-9}
\rule{0mm}{5mm}& ${\cal O}(\alpha_s)$ & ${\cal O}(\alpha_s^2)$ & &
w/o QCD & ${\cal O}(\alpha_s)$ OS & ${\cal O}(\alpha_s)$ $\overline{\rm MS}$ &
${\cal O}(\alpha_s^2)$ OS & ${\cal O}(\alpha_s^2)$ $\overline{\rm MS}$ \\
\hline
160 & 153.1 & 151.5 &   60 & 80.398 & 80.343 & 80.348 & 80.332 & 80.335 \\
& & &  300 & 80.295 & 80.240 & 80.245 & 80.229 & 80.232 \\
& & & 1000 & 80.199 & 80.143 & 80.150 & 80.133 & 80.137 \\
170 & 162.7 & 161.1 &   60 & 80.468 & 80.407 & 80.413 & 80.396 & 80.398 \\
& & &  300 & 80.362 & 80.302 & 80.309 & 80.290 & 80.294 \\
& & & 1000 & 80.265 & 80.204 & 80.212 & 80.193 & 80.198 \\
180 & 172.3 & 170.6 &   60 & 80.540 & 80.474 & 80.480 & 80.462 & 80.464 \\
& & &  300 & 80.433 & 80.367 & 80.374 & 80.354 & 80.359 \\
& & & 1000 & 80.334 & 80.268 & 80.276 & 80.255 & 80.261 \\
190 & 182.0 & 180.2 &   60 & 80.616 & 80.544 & 80.550 & 80.530 & 80.534 \\
& & &  300 & 80.506 & 80.434 & 80.443 & 80.420 & 80.426 \\
& & & 1000 & 80.405 & 80.333 & 80.343 & 80.319 & 80.327 \\
200 & 191.6 & 189.7 &   60 & 80.695 & 80.616 & 80.624 & 80.602 & 80.605 \\
& & &  300 & 80.582 & 80.504 & 80.513 & 80.489 & 80.496 \\
& & & 1000 & 80.479 & 80.400 & 80.412 & 80.386 & 80.395 \\
210 & 201.2 & 199.3 &   60 & 80.777 & 80.692 & 80.700 & 80.676 & 80.680 \\
& & &  300 & 80.661 & 80.576 & 80.587 & 80.560 & 80.568 \\
& & & 1000 & 80.555 & 80.470 & 80.483 & 80.454 & 80.465 \\
\hline\hline
\end{tabular}
\end{table}

In Table~\ref{tab:scheme}, we investigate the scheme dependence of the $M_W$
prediction via $\Delta r$ to ${\cal O}(\alpha\alpha_s)$ and
${\cal O}(\alpha\alpha_s^2)$ within the ranges 160~GeV${}\le M_t\le210$~GeV
and 60~GeV${}\le M_H\le1$~TeV.
For the time being, we suppress the logarithmic terms of RG origin by choosing
$\xi_c=\xi_m=1$.
For this choice, the $\overline{\mbox{MS}}$ values are always in excess of the
the respective OS numbers, {\it i.e.}, the QCD corrections are less negative
in the $\overline{\mbox{MS}}$ scheme.
In Fig.~\ref{fig:scheme}, we display this excess as a function of $M_H$ for
$M_t=168$, 180, and 192~GeV \cite{abe} both in ${\cal O}(\alpha\alpha_s)$ and
${\cal O}(\alpha\alpha_s^2)$.
We observe that the scheme dependence increases both with $M_t$ and $M_H$.
For the $M_t$ and $M_H$ intervals considered in Table~\ref{tab:scheme}, it
ranges from 4.7 to 13.2~MeV in ${\cal O}(\alpha\alpha_s)$ and from 2.2 to
10.3~MeV in ${\cal O}(\alpha\alpha_s^2)$.
As expected, it is significantly reduced as we pass from
${\cal O}(\alpha\alpha_s)$ to ${\cal O}(\alpha\alpha_s^2)$.
For $M_t=180$~GeV, the reduction amounts to 51\% (28\%) for $M_H=60$~GeV
(1~TeV).
In Table~\ref{tab:scheme}, we also list the values of $\bar\mu_t$ in
${\cal O}(\alpha_s)$ and ${\cal O}(\alpha_s^2)$.
In ${\cal O}(\alpha_s^2)$, they lie by a roughly constant amount of 10~GeV
below the respective $M_t$ values.
This may also be seen from Eq.~(\ref{muos}), where, in the case of top, the
coefficients of $\alpha_s(M_t)/\pi$ and $[\alpha_s(M_t)/\pi]^2$ are $-4/3$ and
$-6.458\,784$, respectively.
To assess the significance of the scheme and scale dependences relative to
the overall QCD effect on $M_W$, we also include in Table~\ref{tab:scheme} the
respective $M_W$ values evaluated with the (perturbative) QCD corrections
switched off.
Of course, we cannot reliably eliminate the intrinsic non-perturbative QCD
corrections contained in $\Delta\alpha_{had}^{(5)}$, {\it i.e.}, without
introducing a significant dependence on the poorly-known light-quark masses.

\begin{table}[ht]
\caption{
$M_W$ (in GeV) predicted for $M_t=180$~GeV and $M_H=300$~GeV via $\Delta r$ to
${\cal O}(\alpha\alpha_s)$ and ${\cal O}(\alpha\alpha_s^2)$ in the OS scheme
with $\xi_c$ variable.
}\label{tab:os}
\medskip
\begin{tabular}{|c|c|c|} \hline\hline
\rule{0mm}{5mm}$\xi_c$ & \multicolumn{2}{c|}{$M_W$ [GeV]} \\ \cline{2-3}
\rule{0mm}{5mm}& ${\cal O}(\alpha_s)$ & ${\cal O}(\alpha_s^2)$ \\
\hline
1/8 & 80.346 & 80.351 \\
1/4 & 80.354 & 80.351 \\
1/2 & 80.361 & 80.352 \\
1 & 80.367 & 80.354 \\
2 & 80.372 & 80.357 \\
4 & 80.376 & 80.359 \\
8 & 80.379 & 80.362 \\
\hline\hline
\end{tabular}
\end{table}

In the remainder of this section, we shall stick to the central values
$M_t=180$~GeV and $M_H=300$~GeV, and consider scale variations with
$1/8\le\xi_c,\xi_m\le8$.
In Table~\ref{tab:os} and Fig.~\ref{fig:os}, we study the $\mu_c$ dependence of
$M_W$ to ${\cal O}(\alpha\alpha_s)$ and ${\cal O}(\alpha\alpha_s^2)$ in the OS
scheme.
Obviously, the scale dependence is dramatically reduced, from 33.7~MeV to
11.1~MeV, {\it i.e.} by 67\%, if the ${\cal O}(\alpha\alpha_s^2)$ correction
is taken into account.
This should be compared with the shifts in $M_W$ induced by the QCD
corrections to ${\cal O}(\alpha\alpha_s)$ and ${\cal O}(\alpha\alpha_s^2)$ for
$\mu_c=M_t$, which are $-66.1$~MeV and $-78.6$~MeV, respectively, as may be
seen from Table~\ref{tab:scheme}.
It is interesting to note that 2.4~MeV, {\it i.e.}, 19\%, of the difference
between the ${\cal O}(\alpha\alpha_s)$ and ${\cal O}(\alpha\alpha_s^2)$
evaluations of $M_W$ is due to the three-loop correction to $\Delta r_{rem}$
\cite{che}.
The ${\cal O}(\alpha\alpha_s^2)$ evaluation exhibits a local minimum at
$\xi_c=0.176$.
This is the point advocated by the principle of minimal sensitivity (PMS)
\cite{pms}.
The ${\cal O}(\alpha\alpha_s)$ and ${\cal O}(\alpha\alpha_s^2)$ curves cross
over at $\xi_c=0.183$, the point of fastest apparent convergence (FAC)
\cite{fac}.
In the OS analysis of $\Delta\rho$ to ${\cal O}(\alpha_s^2G_FM_t^2)$ with
$n_f=5$, these points occur at $\xi_c=0.224$ and $\xi_c=0.264$, respectively
\cite{phi}.
We note in passing that the application of the Brodsky--Lepage--Mackenzie (BLM)
\cite{blm} scale-setting criterion to $\Delta\rho$ leads to $\xi_c=0.154$
\cite{avd,phi,vol}, which had been anticipated in the pioneering work of
Ref.~\cite{vol} prior to the advent of the ${\cal O}(\alpha_s^2G_FM_t^2)$
calculation of $\Delta\rho$ \cite{avd}.

\begin{table}[ht]
\caption{
$M_W$ (in GeV) predicted for $M_t=180$~GeV and $M_H=300$~GeV via $\Delta r$ to
${\cal O}(\alpha\alpha_s)$ (upper entries) and ${\cal O}(\alpha\alpha_s^2)$
(lower entries) in the $\overline{\mbox{MS}}$ scheme with $\xi_c$ and $\xi_m$
variable.
}\label{tab:msb}
\medskip
\begin{tabular}{|c|c|c|c|c|c|c|c|} \hline\hline
$\xi_c\backslash\xi_m$ & 1/8 & 1/4 & 1/2 & 1 & 2 & 4 & 8 \\
\hline
1/8 & 80.256 & 80.304 & 80.343 & 80.374 & 80.399 & 80.421 & 80.441 \\
& 80.365 & 80.360 & 80.357 & 80.357 & 80.357 & 80.357 & 80.357 \\
\hline
1/4 & 80.290 & 80.325 & 80.352 & 80.374 & 80.392 & 80.407 & 80.421 \\
& 80.349 & 80.352 & 80.355 & 80.358 & 80.361 & 80.363 & 80.366 \\
\hline
1/2 & 80.318 & 80.342 & 80.360 & 80.374 & 80.386 & 80.395 & 80.404 \\
& 80.343 & 80.349 & 80.354 & 80.358 & 80.362 & 80.366 & 80.369 \\
\hline
1 & 80.342 & 80.356 & 80.366 & 80.374 & 80.380 & 80.386 & 80.390 \\
& 80.344 & 80.350 & 80.355 & 80.359 & 80.363 & 80.366 & 80.368 \\
\hline
2 & 80.362 & 80.368 & 80.372 & 80.374 & 80.376 & 80.377 & 80.378 \\
& 80.347 & 80.352 & 80.356 & 80.359 & 80.362 & 80.364 & 80.366 \\
\hline
4 & 80.379 & 80.378 & 80.377 & 80.375 & 80.372 & 80.370 & 80.369 \\
& 80.353 & 80.356 & 80.358 & 80.360 & 80.361 & 80.362 & 80.363 \\
\hline
8 & 80.394 & 80.387 & 80.381 & 80.375 & 80.369 & 80.364 & 80.360 \\
& 80.360 & 80.360 & 80.360 & 80.360 & 80.360 & 80.360 & 80.359 \\
\hline\hline
\end{tabular}
\end{table}

In Table~\ref{tab:msb} and Fig.~\ref{fig:msb}, we investigate how the
${\cal O}(\alpha\alpha_s)$ and ${\cal O}(\alpha\alpha_s^2)$ calculations of
$M_W$ in the $\overline{\mbox{MS}}$ scheme depend on $\mu_c$ and $\mu_m$.
We notice that the $\mu_c$ dependence is rather feeble for
$\mu_m\approx\bar\mu_t$.
This may be understood by observing that the coefficient of
$\alpha_s(\mu_c)/\pi$ in the QCD expansion of $\Delta\bar\rho$ in
Eq.~(\ref{rhomsb}) is then greatly suppressed \cite{jeg}.
In Fig.~\ref{fig:msb}, the points of minimal sensitivity, {\it i.e.}, with
zero tangents, are marked with ``$x$.''
They are saddle points and gathered in a small strip around $\mu_m=\bar\mu_t$.
Their $(\xi_c,\xi_m)$ coordinates are $(2.732,1.024)$ in
${\cal O}(\alpha\alpha_s)$ and $(0.133,0.853)$ and $(7.127,1.147)$ in
${\cal O}(\alpha\alpha_s^2)$.
For fixed $\xi_c<1$, the ${\cal O}(\alpha\alpha_s)$ value of $M_W$ varies quite
strongly with $\xi_m$, by 185.4~MeV for $\xi_c=1/8$.
This has to be compared with the shift in $M_W$ due to the
${\cal O}(\alpha\alpha_s)$ correction for $\xi_c=\xi_m=1$, which only is
58.6~MeV in size (see Table~\ref{tab:scheme}).
Of course, such an extreme variation cannot be interpreted as the uncertainty
due to the neglect of higher-order QCD corrections.
This rather tells us that our choice of the $\xi_m$ interval width is not
judicious in this case.
Fortunately, the ${\cal O}(\alpha\alpha_s^2)$ calculation is much more stable
under scale variations.
Here, the overall fluctuation is just 25.7~MeV, while the QCD-induced shift in
$M_W$ for $\xi_c=\xi_m=1$ is 73.8~MeV in magnitude (see
Table~\ref{tab:scheme}).
However, this variation is still approximately 2.3 times as large as the one in
the corresponding OS calculation.

\begin{table}[ht]
\caption{
Maximum deviations (in MeV) of the $M_W$ values predicted for $M_t=180$~GeV and
$M_H=300$~GeV via $\Delta r$ to ${\cal O}(\alpha\alpha_s)$ and
${\cal O}(\alpha\alpha_s^2)$ in the OS and $\overline{\mbox{MS}}$ schemes with
$1/\xi_{max}\le\xi_c,\xi_m\le\xi_{max}$ from the respective values for
$\xi_c=\xi_m=1$.
}\label{tab:dev}
\medskip
\begin{tabular}{|c|c|c|c|c|} \hline\hline
\rule{0mm}{5mm}$\log_2\xi_{max}$ & \multicolumn{4}{c|}{$\Delta M_W$ [MeV]} \\
\cline{2-5}
\rule{0mm}{5mm}& ${\cal O}(\alpha_s)$ OS &
${\cal O}(\alpha_s)$ $\overline{\rm MS}$ & ${\cal O}(\alpha_s^2)$ OS &
${\cal O}(\alpha_s^2)$ $\overline{\rm MS}$ \\
\hline
1/2 & $+2.5$ & $+4.5$ & $+1.2$ & $+1.9$ \\
    & $-2.8$ & $-5.2$ & $-1.1$ & $-2.4$ \\
\hline
1   & $+4.9$ & $+11.4$ & $+2.4$ & $+3.6$ \\
    & $-5.8$ & $-14.4$ & $-2.1$ & $-5.0$ \\
\hline
3/2 & $+7.0$ & $+20.8$ & $+3.7$ & $+5.3$ \\
    & $-9.0$ & $-28.6$ & $-2.9$ & $-7.4$ \\
\hline
2   & $+9.1$ & $+32.9$ & $+5.0$ & $+6.9$ \\
    & $-12.6$ & $-49.4$ & $-3.4$ & $-10.0$ \\
\hline
5/2 & $+10.9$ & $+48.0$ & $+6.2$ & $+8.4$ \\
    & $-16.6$ & $-78.6$ & $-3.6$ & $-12.8$ \\
\hline
3   & $+12.7$ & $+66.5$ & $+7.5$ & $+9.8$ \\
    & $-21.0$ & $-118.8$ & $-3.6$ & $-16.0$ \\
\hline\hline
\end{tabular}
\end{table}

Since it is hard to extract precise numbers from the contour plots in
Fig.~\ref{fig:msb}, we list in Table~\ref{tab:dev} the maximum deviations
of the $\overline{\mbox{MS}}$ evaluations of $M_W$ to
${\cal O}(\alpha\alpha_s)$ and ${\cal O}(\alpha\alpha_s^2)$ within the variable
range $1/\xi_{max}\le\xi_c,\xi_m\le\xi_{max}$ from the respective values for
$\xi_c=\xi_m=1$.
For completeness, we also list the corresponding numbers for the OS analyses to
${\cal O}(\alpha\alpha_s)$ and ${\cal O}(\alpha\alpha_s^2)$ of
Fig.~\ref{fig:os}.
It is interesting to observe that, before the appearance of the
${\cal O}(\alpha\alpha_s^2)$ corrections in Eq.~(\ref{dr}) \cite{avd,che}, the
uncertainty due to the lack of these terms could have been estimated from the
scale variation of the ${\cal O}(\alpha\alpha_s)$ OS ($\overline{\mbox{MS}}$)
calculation with $\xi_{max}=4$ (2.1).
Thus, we may expect that similar scale variations will also yield meaningful
results in the next order.

The separation of $\mu_c$ and $\mu_m$ is perhaps not so easy to motivate on
physical grounds.
In Fig.~\ref{fig:msbi}, we analyze the scale dependence of the
$\overline{\mbox{MS}}$ calculation of $M_W$ to ${\cal O}(\alpha\alpha_s)$ and
${\cal O}(\alpha\alpha_s^2)$ identifying $\mu_c=\mu_m=\xi\bar\mu_t$.
We observe that the ${\cal O}(\alpha\alpha_s)$ analysis is very unstable for
$\xi\ll1$.
For $\xi=1/8$, we have $M_W=80.256$~GeV (see Table~\ref{tab:msb}), which is
way below the $M_W$ range considered in Fig.~\ref{fig:msbi}.
On the other hand, the ${\cal O}(\alpha\alpha_s^2)$ curve nicely oscillates
around the $M_W$ value at $\xi=1$, with a band-width of 13.5~MeV.
In this one-dimensional analysis, the PMS points appear at $\xi=1.601$ in
${\cal O}(\alpha\alpha_s)$ and at $\xi=0.280$ and $\xi=2.898$
in ${\cal O}(\alpha\alpha_s^2)$.
They are indicated by ``$o$'' in the contour plots of Fig.~\ref{fig:msb}.
There is one point of FAC within Fig.~\ref{fig:msbi}, at $\xi=0.413$.
Another one is located at $\xi=8.358$.
The BLM criterion only applies to $\mu_c$, but not to $\mu_m$, so that its
implementation in our $\overline{\mbox{MS}}$ analysis is ambiguous \cite{mac}.

\section{Conclusions}

In this paper, we have extended an existing calculation of $\Delta r$ in the
OS scheme \cite{sto,fan} to next-to-leading order in QCD by incorporating new
three-loop results \cite{avd,che}.
We have then converted this analysis to the $\overline{\mbox{MS}}$ scheme of
quark-mass renormalization in QCD.
Armed with these results, we have analyzed the scheme and scale dependences
of the $M_W$ value predicted for given values of $M_t$ (pole mass) and $M_H$.
We have verified that both scheme and scale dependences are considerably
reduced if the next-to-leading-order QCD corrections are taken into account.
For $M_t=180$~GeV and $M_H=300$~GeV, the scheme dependence at the central
renormalization point, $\xi_c=\xi_m=1$, is reduced from 7.5~MeV to 4.8~MeV
(see Table~\ref{tab:scheme} and Fig.~\ref{fig:scheme}).
The scale dependences within the interval $1/4\le\xi_c,\xi_m\le4$ are decreased
from 21.7~MeV to 8.4~MeV in the OS scheme and from 82.2~MeV to 16.9~MeV in the
$\overline{\mbox{MS}}$ scheme (see Tables~\ref{tab:os}, \ref{tab:msb},
\ref{tab:dev} and Figs.~\ref{fig:os}, \ref{fig:msb}).

Although the numbers presented here are uniquely determined by the generally
accepted rules of perturbation theory, their interpretation in terms of a
QCD-related theoretical uncertainty in $M_W$ allows for a certain amount
freedom.
Nevertheless, we shall propose an algorithm to extract a central value and an
error of $M_W$.
If we adopt the point of view that the OS and $\overline{\mbox{MS}}$ analyses
represent two independent theoretical determinations of $M_W$ with their
individual errors, then we may combine them assuming Gaussian statistics as we
usually do with independent experimental measurements \cite{pdg}.
It is plausible and conservative to identify the error in each scheme with the
absolute of the maximum deviation of $M_W$ in the interval
$1/\xi_{max}\le\xi_c,\xi_m\le\xi_{max}$ from the central value, at
$\xi_c=\xi_m=1$.
This will leave room for the reader to select his preferred value of
$\xi_{max}$.
For illustration, we shall again assume that $M_t=180$~GeV and $M_H=300$~GeV.
The OS and $\overline{\mbox{MS}}$ central values of $M_W$ and the respective
errors may then be read off from Tables~\ref{tab:os}, \ref{tab:msb}, and
\ref{tab:dev}.
For $\xi_{max}=4$, we so obtain $(80.367\pm0.012)$~GeV to
${\cal O}(\alpha\alpha_s)$ and $(80.355\pm0.004)$~GeV to
${\cal O}(\alpha\alpha_s^2)$.
We would like to point out that the ${\cal O}(\alpha\alpha_s^2)$ central value
is encompassed by the ${\cal O}(\alpha\alpha_s)$ error, and that, in each
order, the OS and $\overline{\mbox{MS}}$ central values lie within the error.
This reassures us of the soundness of our procedure.
The corresponding results for other values of $\xi_{max}$ may be seen from
Fig.~\ref{fig:err}.
It is clear that this approach does not make sense if $\xi_{max}$ is so small
that the scale variations of the OS and $\overline{\mbox{MS}}$ calculations do
not overlap.

Apart from the theoretical uncertainty due to the lack of knowledge of QCD
corrections to $\Delta r$ beyond three loops, which we have estimated here,
there is another source of QCD-related error, namely the one connected with
$\Delta\alpha_{had}^{(5)}$, which is extracted from experimental data of
$e^+e^-\to hadrons$ via a dispersion relation \cite{eid,swa}.
In Ref.~\cite{eid}, this error has been estimated to be $\pm0.0007$, which,
for $M_t=(180\pm12)$~GeV and 60~GeV${}\le M_H\le{}$1~TeV, translates into an
error of approximately $\pm13$~MeV in $M_W$.
Other recent determinations of $\Delta\alpha_{had}^{(5)}$ \cite{swa} agree
with the result of Ref.~\cite{eid} within less than two standard deviations of
the latter.

Finally, we would like to mention that the lack of knowledge of three-loop and
subleading two-loop electroweak corrections to $\Delta r$ represents another
source of theoretical error on the $M_W$ determination from the measured muon
lifetime.
The magnitude of the recently calculated ${\cal O}(G_F^2M_t^2M_W^2)$
correction to $\Delta\rho$ \cite{deg} indicates that the still uncontrolled
${\cal O}(G_F^2M_t^2M_W^2)$ term of $\Delta r$ might have the potential to
jeopardize the accuracy of the indirect $M_W$ determination.
The study of the scheme and scale dependences of the dominant two-loop
electroweak corrections provides us with a clue to the size of higher-order
electroweak effects \cite{boc}.

\bigskip
\centerline{\bf ACKNOWLEDGEMENTS}
\smallskip\noindent
I am grateful to Bill Bardeen for carefully reading the manuscript, to Gerhard
Buchalla for a useful comment on the separation of coupling and mass
renormalization scales, to Kostja Chetyrkin for making available to me the
results of Ref.~\cite{che} in algebraic form, to Paul Mackenzie for a helpful
conversation about Ref.~\cite{blm}, to Sergei Larin and Levan Surguladze for
clarifying communications regarding the decoupling relation \cite{lar}, and to
Alberto Sirlin for illuminating discussions concerning the scheme and scale
dependences of $\Delta\rho$.
I am indebted to the FNAL Theory Group for inviting me as a Guest Scientist and
for the great hospitality extended to me.

\appendix
\section{Appendix}

In this Appendix, we shall provide a few general relations which are useful
for implementing the $\mu$ dependence of the QCD coupling constant and the
quark masses in the $\overline{\mbox{MS}}$ scheme.
We shall take the colour gauge group to be SU($N_c$);
$C_F=(N_c^2-1)/(2N_c)$ and $C_A=N_c$ are the Casimir operators of its
fundamental and adjoint representations, respectively, and $T_F=1/2$ is
the trace normalization of its fundamental representation.
We shall keep the number of active quark flavours, $n_f$, arbitrary.
The RG equations for the so-called couplant, $a(\mu)=\alpha_s(\mu)/\pi$, and
$m(\mu)$ may be found, {\it e.g.}, in Ref.~\cite{lar}.
For the reader's convenience, we list them here:
\begin{eqnarray}
\label{rgea}
{da\over d\ln\mu}&\n=\n&\beta(a)
=-a^2\left[\beta_0+\beta_1a+\beta_2a^2+{\cal O}(a^3)\right],\\
\label{rgem}
{d\ln m\over d\ln\mu}&\n=\n&-\gamma_m(a)
=-a\left[\gamma_0+\gamma_1a+\gamma_2a^2+{\cal O}(a^3)\right],
\end{eqnarray}
where \cite{gro}
\begin{eqnarray}
\label{beta}
\beta_0&\n=\n&{1\over4}\left({11\over3}C_A-{4\over3}T_Fn_f\right),
\nonumber\\
\beta_1&\n=\n&{1\over16}\left({34\over3}C_A^2-4C_FT_Fn_f
-{20\over3}C_AT_Fn_f\right),
\nonumber\\
\beta_2&\n=\n&{1\over64}\left({2857\over54}C_A^3+2C_F^2T_Fn_f
-{205\over9}C_FC_AT_Fn_f-{1415\over27}C_A^2T_Fn_f+{44\over9}C_FT_F^2n_f^2
\right.\nonumber\\
&\n\n&\left.+{158\over27}C_AT_F^2n_f^2\right)
\end{eqnarray}
are the first three coefficients of the Callan-Symanzik beta function and
\cite{tar,nac}
\begin{eqnarray}
\label{ano}
\gamma_0&\n=\n&{3\over4}C_F,
\nonumber\\
\gamma_1&\n=\n&{1\over16}\left({3\over2}C_F^2+{97\over6}C_FC_A
-{10\over3}C_FT_Fn_f\right),
\nonumber\\
\gamma_2&\n=\n&{1\over64}\left\{{129\over2}C_F^3-{129\over4}C_F^2C_A
+{11413\over108}C_FC_A^2+C_F^2T_Fn_f[48\zeta(3)-46]
\right.\nonumber\\
&\n\n&\left.+C_FC_AT_Fn_f\left[-48\zeta(3)-{556\over27}\right]
-{140\over27}C_FT_F^2n_f^2\right\}
\end{eqnarray}
are the first three coefficients of the quark-mass anomalous dimension.
Here, $\zeta$ is Riemann's zeta function, with value
$\zeta(3)\approx1.202\,057$.
Our aim is to evaluate $a=a(\mu)$ and $m=m(\mu)$ for $\mu$ arbitrary, assuming
that their values, $a_0=a(\mu_0)$ and $m_0=m(\mu_0)$, are known at some
starting scale, $\mu_0$.

Beyond the leading order, Eq.~(\ref{rgea}) cannot be exactly solved for $a$.
A perturbative solution reads
\begin{equation}
\label{expa}
a=a_0\left[1-a_0\beta_0\ell+a_0^2\ell(\beta_0^2\ell-\beta_1)
+a_0^3\ell\left(-\beta_0^3\ell^2+{5\over2}\beta_0\beta_1\ell-\beta_2\right)
+{\cal O}(a_0^4\ell^4)\right],
\end{equation}
where $\ell=\ln(\mu^2/\mu_0^2)$.
The leading logarithms in Eq.~(\ref{expa}) may be resummed by writing
\cite{gil}
\begin{equation}
\label{resa}
a={a_0\over1+a_0\ell\left[\beta_0+a_0\beta_1
+a_0^2(-\beta_0\beta_1\ell/2+\beta_2)\right]}.
\end{equation}
We recover the exact leading-order solution of Eq.~(\ref{rgea}) by discarding
the terms of ${\cal O}(a_0^2)$ in the denominator of Eq.~(\ref{resa}).

Knowing the $\mu$ dependence of $a$, it is sufficient to obtain $m$ as a
function of $a$.
Dividing Eq.~(\ref{rgem}) by Eq.~(\ref{rgea}), we obtain a differential
equation which may be exactly solved for $m$, given the coefficients of
$\beta$ and $\gamma_m$ to a certain order.
The exact solutions to leading and next-to-leading orders read
\begin{eqnarray}
\label{onem}
m&\n=\n&m_0\left({a\over a_0}\right)^{\gamma_0/\beta_0},\\
\label{twom}
m&\n=\n&m_0\left({a\over a_0}\right)^{\gamma_0/\beta_0}
\left({\beta_0+a\beta_1\over\beta_0+a_0\beta_1}\right)^{\gamma_1/\beta_1
-\gamma_0/\beta_0},
\end{eqnarray}
respectively.
A perturbative solution of Eq.~(\ref{rgem}) similar to Eq.~(\ref{expa}) is
given by
\begin{eqnarray}
\label{expm}
m&\n=\n&m_0\left\{1-a_0\gamma_0\ell
+a_0^2\ell\left[{\gamma_0\over2}(\beta_0+\gamma_0)\ell-\gamma_1\right]
+a_0^3\ell\left[-\gamma_0\left({\beta_0^2\over3}+{\beta_0\gamma_0\over2}
+{\gamma_0^2\over6}\right)\ell^2
\right.\right.\nonumber\\
&\n\n&\left.\left.
+\left(\beta_0\gamma_1+{\beta_1\gamma_0\over2}+\gamma_0\gamma_1\right)\ell
-\gamma_2\right]+{\cal O}(a_0^4\ell^4)\right\},
\end{eqnarray}
where $\ell$ is defined below Eq.~(\ref{expa}).
By iterating Eq.~(\ref{expm}), we may generate a closed expression for the mass
parameter $\bar\mu=m(\bar\mu)$ in terms of $a$ and $m$ for $\mu$ arbitrary:
\begin{eqnarray}
\label{expmu}
\bar\mu&\n=\n&m\left\{1+a\gamma_0l
+a^2l\left[{\gamma_0\over2}(\beta_0+\gamma_0)l-2\gamma_0^2+\gamma_1\right]
+a^3l\left[\gamma_0\left({\beta_0^2\over3}+{\beta_0\gamma_0\over2}
+{\gamma_0^2\over6}\right)l^2
\right.\right.\nonumber\\
&\n\n&\left.\left.
+\left(-3\beta_0\gamma_0^2+\beta_0\gamma_1+{\beta_1\gamma_0\over2}-2\gamma_0^3
+\gamma_0\gamma_1\right)l
+4\gamma_0^3-4\gamma_0\gamma_1+\gamma_2\right]+{\cal O}(a^4l^4)\right\},
\end{eqnarray}
where $\l=\ln(\mu^2/m^2)$.
The right-hand side of Eq.~(\ref{expmu}) is manifestly RG invariant through
${\cal O}(a^3)$.

In the remainder of this section, we shall consider QCD with one massive quark
and $n_f-1$ massless flavours.
The relation between the pole mass, $M$, and $m(M)$ at next-to-leading order
is given by \cite{tar,gra}
\begin{equation}
\label{pole}
{M\over m(M)}=1+a(M)C_F+a^2(M)K_0+{\cal O}(a^3),
\end{equation}
where \cite{gra}
\begin{eqnarray}
\label{kzero}
K_0&\n=\n&C_F\left[{3\over4}\zeta(2)-{3\over8}\right]
+C_F^2\left[{3\over4}\zeta(3)+\zeta(2)\left(-3\ln2+{15\over8}\right)
+{121\over128}\right]
\nonumber\\
&\n\n&
+C_FC_A\left[-{3\over8}\zeta(3)+\zeta(2)\left({3\over2}\ln2-{1\over2}\right)
+{1111\over384}\right]
+C_Fn_f\left[-{\zeta(2)\over4}-{71\over192}\right]
\nonumber\\
&\n\approx\n&17.151\,430-1.041\,367\,n_f.
\end{eqnarray}
Here, $\zeta(2)=\pi^2/6$ and the value of $\zeta(3)$ is listed below
Eq.~(\ref{ano}).
Using Eq.~(\ref{pole}) along with Eqs.~(\ref{expa}) and (\ref{expm}), we may
express $M$ entirely in terms of $\overline{\mbox{MS}}$ parameters, viz.\
\begin{eqnarray}
\label{osms}
M&\n=\n&m\left\{1+a(\gamma_0l+C_F)
+a^2\left[{\gamma_0\over2}(\beta_0+\gamma_0)l^2
+\left(-2\gamma_0^2+\gamma_1+C_F(\beta_0+\gamma_0)\right)l-2\gamma_0C_F
\right.\right.\nonumber\\
&\n\n&\left.\left.\vphantom{\gamma_0\over2}
+K_0\right]+{\cal O}(a^3l^3)\right\},
\end{eqnarray}
where $l$ is defined below Eq.~(\ref{expmu}).
The right-hand side of Eq.~(\ref{osms}) is RG invariant
through ${\cal O}(a^2)$ as it must be.
Substituting Eq.~(\ref{pole}) into Eq.~(\ref{expmu}) evaluated at $\mu=M$,
we obtain a closed expression of $\bar\mu$ in terms of $M$ \cite{ste},
\begin{equation}
\label{muos}
\bar\mu=M\left\{1-a(M)C_F+a^2(M)\left[C_F(2\gamma_0+C_F)-K_0\right]+
{\cal O}(a^3)\right\}.
\end{equation}
In Eqs.~(\ref{pole}), (\ref{osms}), and (\ref{muos}), $a$ refers to $n_f$ quark
flavours, {\it i.e.}, $a=a^{(n_f)}$.
Finally, we list the matching condition for $a$ at the heavy-flavour
threshold, $\mu=M$.
It follows from the decoupling relation found in Ref.~\cite{lar} and reads
\begin{equation}
\label{mat}
a^{(n_f)}(M)=a^{(n_f-1)}(M)\left\{1+\left[a^{(n_f-1)}(M)\right]^2
T_F\left({15\over16}C_F-{2\over9}C_A\right)+{\cal O}(a^3)\right\}.
\end{equation}

\begin{figure}[p]

\vskip-6cm

\centerline{\bf FIGURE CAPTIONS}

\caption{\label{fig:scheme} Difference (in MeV) between the
$\overline{\mbox{MS}}$ and OS evaluations of $M_W$ via $\Delta r$ to
${\cal O}(\alpha\alpha_s)$ (dashed lines) and ${\cal O}(\alpha\alpha_s^2)$
(solid lines) with $\xi_c=\xi_m=1$ as a function of $M_H$ for
$M_t=168$, 180, and 192~GeV.
Upper curves correspond to larger $M_t$ values.}

\caption{\label{fig:os} $\mu_c$ dependence of the OS evaluation of $M_W$ (in
GeV) via $\Delta r$ to ${\cal O}(\alpha\alpha_s)$ (dashed line) and
${\cal O}(\alpha\alpha_s^2)$ (solid line) for $M_t=180$~GeV and
$M_H=300$~GeV.}

\caption{\label{fig:msb} $\mu_c$ and $\mu_m$ dependences of the
$\overline{\mbox{MS}}$ evaluation of $M_W$ via $\Delta r$ to (a)
${\cal O}(\alpha\alpha_s)$ and (b) ${\cal O}(\alpha\alpha_s^2)$ for
$M_t=180$~GeV and $M_H=300$~GeV.
The contours of constant deviation (in MeV) from the value at $\xi_c=\xi_m=1$
(marked by $+$) are shown in the $(\log_2\xi_c,\log_2\xi_m)$ plane.
The saddle points in this plane are marked with ``$x$;'' the maxima and minima
on the diagonal $\xi_c=\xi_m$ are marked with ``$o$.''}

\caption{\label{fig:msbi} $\mu_c$ dependence of the $\overline{\mbox{MS}}$
evaluation of $M_W$ (in GeV) via $\Delta r$ to ${\cal O}(\alpha\alpha_s)$
(dashed line) and ${\cal O}(\alpha\alpha_s^2)$ (solid line) with $\mu_c=\mu_m$
for $M_t=180$~GeV and $M_H=300$~GeV.}

\caption{\label{fig:err} Central values and QCD-related errors of $M_W$ (in
GeV) evaluated via $\Delta r$ to ${\cal O}(\alpha\alpha_s)$ (dashed lines) and
${\cal O}(\alpha\alpha_s^2)$ (solid lines) with
$1/\xi_{max}\le\xi_c,\xi_m\le\xi_{max}$ for $M_t=180$~GeV and $M_H=300$~GeV
as a function of $\xi_{max}$.}

\end{figure}

\newpage

\begin{figure}[ht]
\epsfig{figure=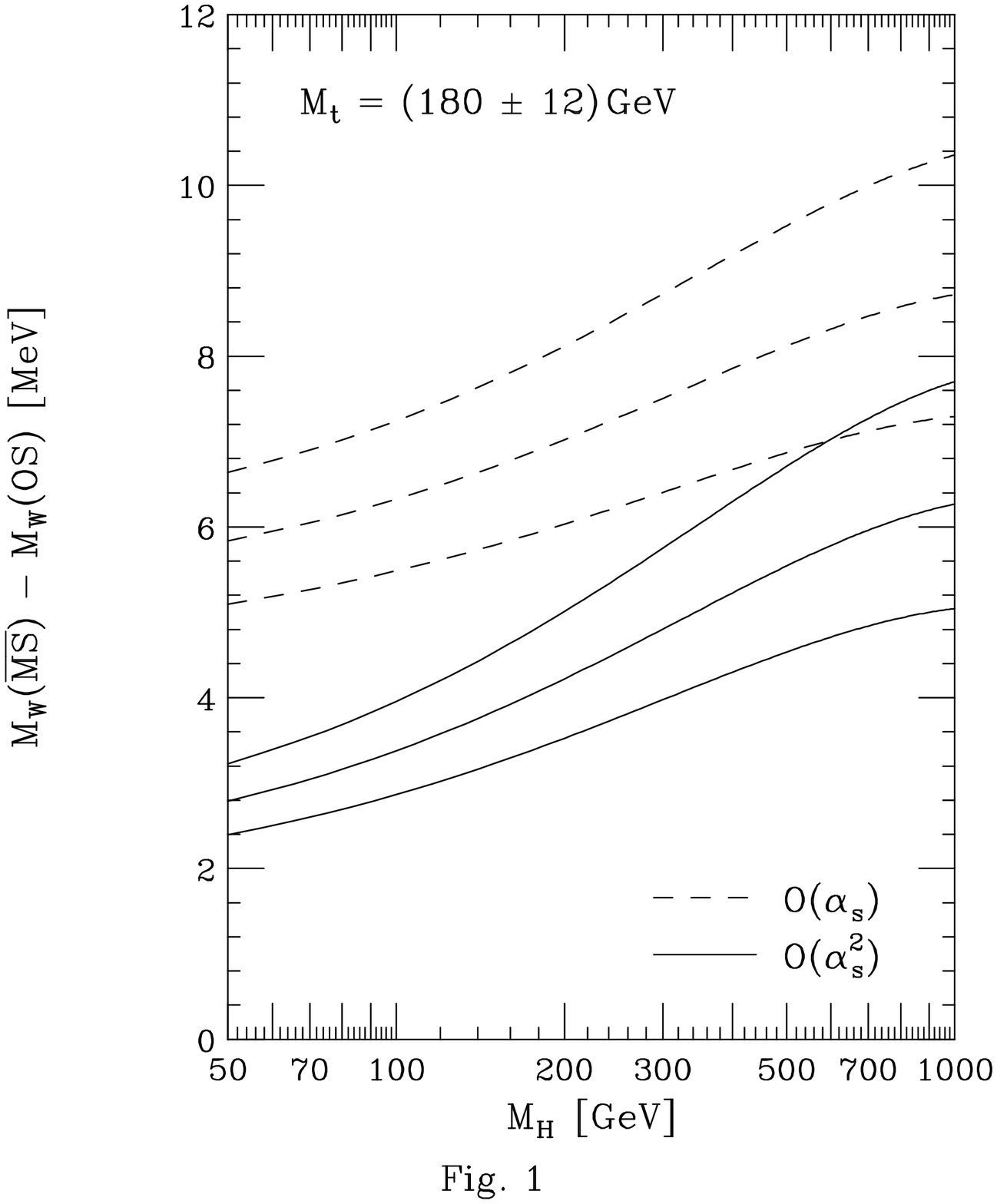,height=\textheight,width=\textwidth}
\end{figure}

\newpage

\begin{figure}[ht]
\epsfig{figure=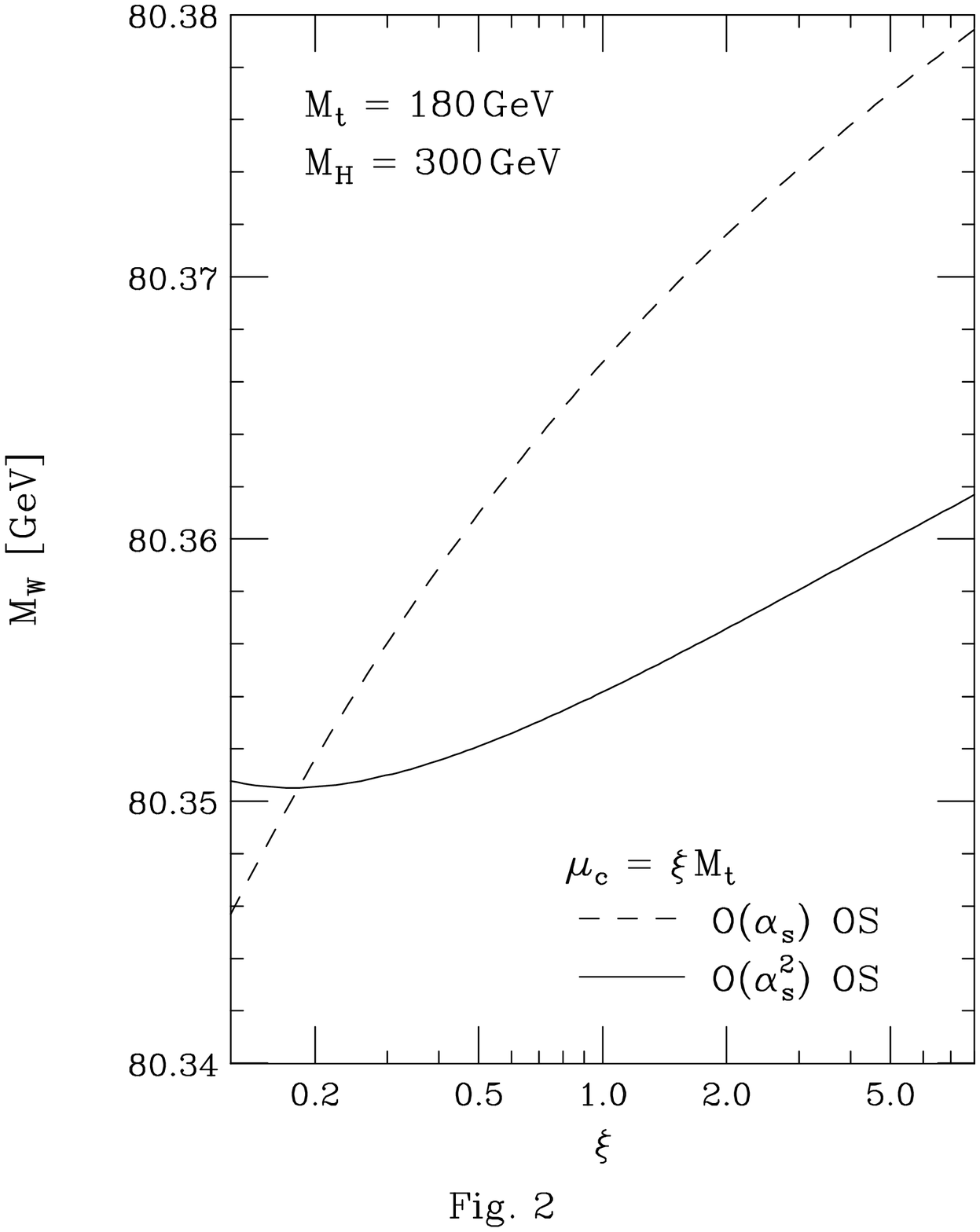,height=\textheight,width=\textwidth}
\end{figure}

\newpage

\begin{figure}[ht]
\epsfig{figure=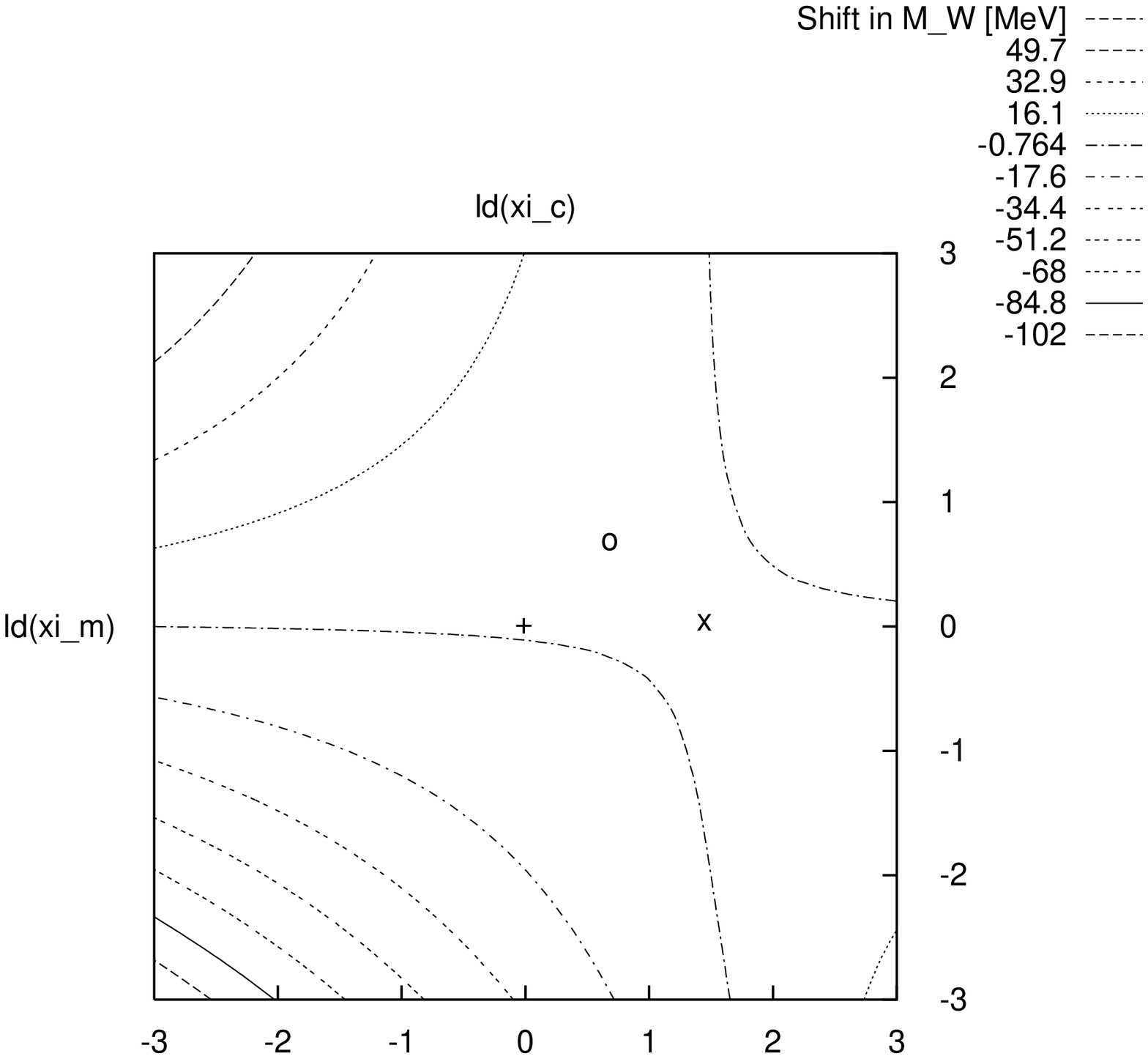,width=\textwidth,angle=-90}
\end{figure}

\newpage

\begin{figure}[ht]
\epsfig{figure=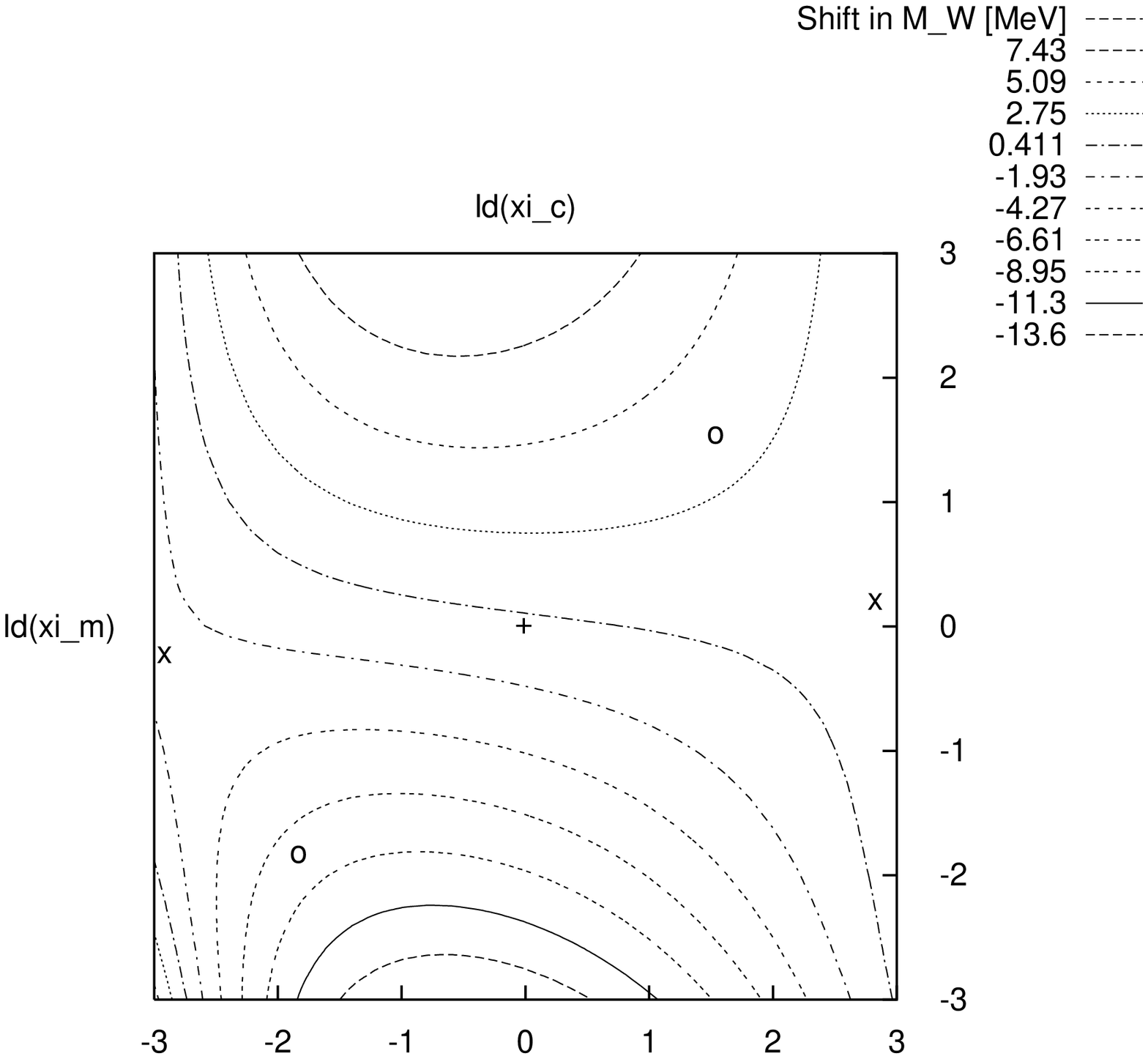,width=\textwidth,angle=-90}
\end{figure}

\newpage

\begin{figure}[ht]
\epsfig{figure=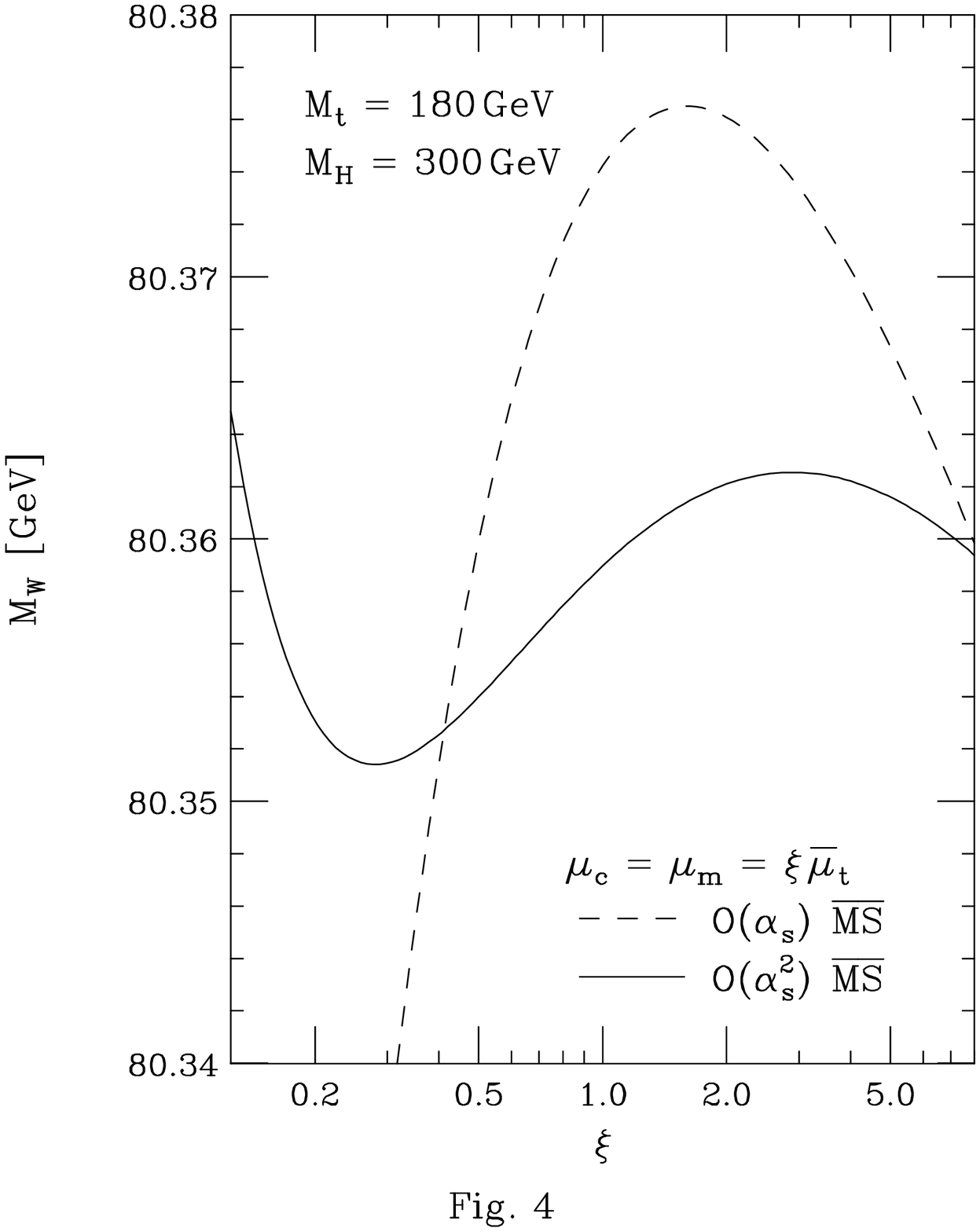,height=\textheight,width=\textwidth}
\end{figure}

\newpage

\begin{figure}[ht]
\epsfig{figure=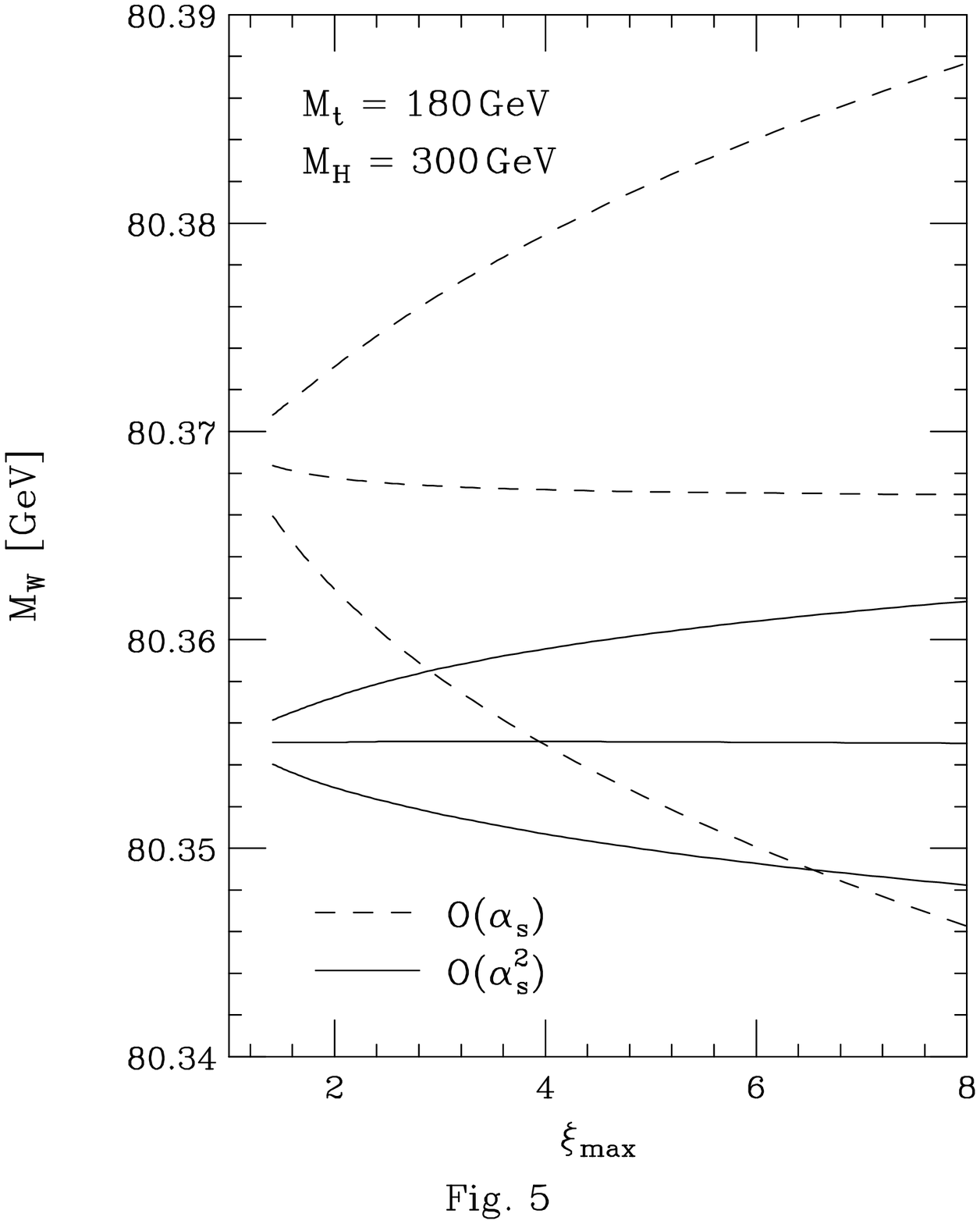,height=\textheight,width=\textwidth}
\end{figure}

\end{document}